\documentstyle[prb,aps,multicol,epsf]{revtex}
\twocolumn
\begin{document}
\draft
\title{
Charge pumping in a quantum wire
driven by a series of local time-periodic potentials
}
\author{Shi-Liang Zhu$^{1,2}$ and 
Z. D. Wang$^{1,3}$ \thanks{To whom correspondence should be addressed.
Email address: zwang@hkucc.hku.hk}
}
\address{ 
$^{1}$Department of Physics, University of Hong Kong, Pokfulam Road,
Hong Kong, China\\
$^{2}$Department of Physics, South China Normal University,
Guangzhou 510631, China\\
$^{3}$ Department of Material Science and Engineering, University of Science and Technology of China, Hefei, China
}
\address{\mbox{}}
\date{Phys. Rev. B 65, 155313 (2002)}
\address{\parbox{14cm}{\rm \mbox{}\mbox{}
We develop a method to calculate
electronic transport properties through a mesoscopic
scattering region in the presence of a series of time-periodic potentials.
Using the method, the quantum charge pumping
driven by time-periodic potentials is studied. 
Jumps in the pumped current
are observed at the peak positions of the Wigner delay time.
Our main results in both the weak pumping 
and strong pumping regimes
are consistent with 
experimental results.
More interestingly, we also observed the
nonzero pumping at the phase difference $\phi=0$ and
addressed its relevance to the experimental result.
}}
\address{\mbox{}}
\address{\parbox{14cm}{\rm PACS numbers: 73.23.-b, 72.10.Bg,
73.50.Pz, 03.65.Bz}}
\maketitle   

\narrowtext
\newpage

A parametric electron pump has attracted considerable attention in recent 
years.~\cite{Thouless,Zhou,Brouwer,Niu,Switkes,Altshuler,Buttiker,Andreev,Aleiner,Levinson,Wagner,Wei}
It is a device that generates a dc current at zero bias by cyclic deformations of 
system parameters.\cite{Thouless,Zhou,Brouwer}
The quantum pumping mechanism 
was originally proposed by Thouless,\cite{Thouless} who
studied the integrated particle current produced by a slow periodic variation of the potential, 
and showed that in a finite torus the integral of the current over a period can vary continuously, but 
 it must
have an integer value in an infinite periodic system with full bands. Such
quantized charge transport
was proposed to
become an electric current standard.~\cite{Niu}

Quite recently, the charge pumping was observed 
experimentally.~\cite{Switkes} For technical reasons, instead of
measuring charge currents, the pumped dc
voltage $V_{dot}$ is measured in a quantum dot where
two gates with oscillating voltages control the deformation
of the shape of the dot.
For weak pumping,
the observed charge pumping has a sinusoidal dependence 
on the phase difference
$\phi$ between the two shape-distorting ac voltages
applied to the gates, and
is proportional to the square of pumping
strength $V$. For strong pumping, the pumped current
deviates from the square dependence on $V$ 
and becomes nonsinusoidal, being always
antisymmetric about $\phi=\pi$.
The charge pumping may have a close relation to
the adiabatic Berry's phase since
the evolution of the system
is cyclic and is controlled
by several system parameters, referred to as
the parametric pumping.
Based on this understanding,
the total charge pumped per cycle is proportional to
the area enclosed by the path
in the parameter space, and
nonzero pumping current requires at least two
parameters. \cite{Switkes,Altshuler}
The pumped charge drived by two parameters should be zero
if two parameters are in phase ($\phi=0$)) since
the area enclosed by the path is zero.
However, it is in contradiction with the observed current
$I(\phi=0)\not= 0$.\cite{Switkes}
One of possible mechanisms of nonzero currents for $\phi=0$ 
is photovoltaic effects introduced in Ref.\cite{Zhou}, where
a surprising result, nonzero dc current generated by a single
pumping gate voltage, is also reported.
The general physics of a quantum pump has been the subject of 
several theoretical analyses.\cite{Zhou,Brouwer} 
Zhou {\it et al} demonstrated that
at low temperatures both the magnitude and the sign of the pumped
charges are sample specific quantities, and the typical value in disordered 
(chaotic) systems turns out to be
determined by quantum interference effects.
Another general expression for the
average transmitted charge current was derived by Brouwer~\cite{Brouwer}
under the adiabatic
condition and based on the time-dependent S-matrix method,\cite{Buttiker} 
which appears to be quite
successful for (adiabatic) weak pumping.
Adiabaticity here means that the oscillating period
$\tau$ of the system is much larger than
the Wigner delay time $\tau_w$.\cite{Brouwer,Andreev}
Note that the adiabatic condition does not simply imply
that the pumping strength $V$ should be very small.
In fact,  the  adiabatic condition requires
that  $\tau$ must be larger as $V$ increases. 
On the other hand, 
the pumping was not weak in the experiments.~\cite{Switkes}
The main purpose of the paper
is to develop a theory, which is also applicable in
the case of strong pumping.
By using the Floquet theorem, the photon-assisted transport
has been taken into account\cite{Wagner}.
We calculate the pumped current
through a mesoscopic region in the presence of time-periodic
potentials.
Our main results in the weak pumping regime,
as well as those in the strong pumping regime
are consistent with the
experiment reported in Ref.\cite{Switkes}.

Consider electrons
transmitting through a one-dimensional scattering region
ranging from $x_0$ to $x_0+\delta$. The potential is
given by
\begin{equation}
\label{potential1}
V(x,t)=\left\{ \begin{array}{cc}
0, & x_0<0,\ x>x_0+\delta  \\
V_s(x,t), & x_0 \leq x \leq x_0+\delta 
\end{array}
\right.
\end{equation}
with $V_s(x,t)=V_0+V_s cos(\omega t+\phi_s)$.
The Schr\"{o}dinger equation can be written as
\begin{equation}
\label{Schrodinger}
i\hbar\frac{\partial \Psi (x,t)}{\partial t}=-\frac{\hbar^2}{2m^*}
\frac{\partial^2 \Psi(x,t)}{\partial x^2}+V(x,t)\Psi(x,t),
\end{equation}
with $m^*$ as the electron effective mass.
Equation \ref{Schrodinger} can be solved by using
the Floquet theorem.\cite{Burmeister} By setting
$\Psi_{Fl}(x,t)=e^{-iE_{Fl}t/\hbar}\psi (x,t),$
where $E_{Fl}$ is the Floquet eigenenergy and $\psi (x,t)$
is a periodic function $\psi (x,t)=\psi(x,t+\tau)$
with period $\tau=2\pi/\omega$, the Schr\"{o}dinger
equation takes the form
$$
E_{Fl}\psi(x,t)=-\frac{\hbar^2}{2m^*}
\frac{\partial^2 \psi(x,t)}{\partial x^2}
-i\hbar\frac{\partial \psi (x,t)}{\partial t}
+V(x,t)\psi(x,t).
$$
Substituting $\psi(x,t)=g(x)f(t)$, we have two separated
equations with an introduced constant $E$,
\begin{eqnarray}
\label{gt}
-\frac{\hbar^2}{2m^*}\frac{\partial^2}{\partial x^2}g(x)
+V_0g(x) &=&  Eg(x), \\
\label{ft}
i\hbar\frac{\partial f(t)}{\partial t}-V_s cos(\omega t +\phi_s)f(t)
&=& (E-E_{Fl})f(t).
\end{eqnarray}
Integrating Eq.(\ref{ft}) gives
\begin{equation}
\label{Solution-ft}
f(t) = e^{\frac{iV_s sin\phi_s}{\hbar\omega}-\frac{i(E-E_{Fl})t}{\hbar}}
          \sum\limits_{\nu=-\infty}^{\infty}
     e^{-\nu \phi_s}J_\nu(\frac{V_s}{\hbar\omega})e^{-i\nu \omega t},
\end{equation}
where $J_\nu (x)$ is the Bessel function of the
first kind of order $\nu$.
Since $f(t)$ is periodic in time
with period $\tau$, it follows from Eq.(\ref{Solution-ft})
that $E-E_{Fl}=m\omega$ with
$m$ as an integer. The equation for $g(x)$ has a solution
\begin{equation}
\label{Solution-gx}
g(x)=e^{\pm ik^s_m x},\ (k^s_m)^2=2m^*(E_{Fl}+m\hbar\omega-V_0)/\hbar^2.
\end{equation}
Thus $\psi (x,t)$ becomes
\begin{equation}
\label{Phi_m}
\psi_m(x,t)=e^{i(V_s/\hbar\omega)sin\phi_s \pm ik^s_m x}
\sum\limits_n F_{n-m} e^{-in\omega t},
\end{equation}
with $F_{n-m}=exp[-i(n-m)\phi_s]J_{n-m}(V_s/\hbar\omega)$.

We consider an incoming wave from the left
with the energy $E_0=\hbar^2 k_0^2/2m^*$, then the
outgoing waves should be
divided into different modes $E_n$, which satisfies
$E_n=E_0+n\hbar\omega$ with
$n=0,\pm1,\pm2,\cdots$. The propagating modes mean that 
$E_n>0$, while the evanescent modes mean that $E_n\leq 0$.
The latter
exists only in the neighborhood of the
oscillating barrier and do not propagate.
Denote $k_n=\sqrt{2m^*E_n}/\hbar$,
the solutions
of the Schr\"{o}dinger equation can be written as
\begin{eqnarray*}
\Psi_l &=& \sum\limits_{n=-\infty}^{\infty} (A_n^i
      e^{ik_n x}+A_n^o e^{-ik_n x})e^{-iE_n t/\hbar},\ (x<x_0)\\
\Psi_s &=& e^{-\frac{iE_{Fl}t}{\hbar}}\sum\limits_{m,n=-\infty}^{\infty}
                (a_{m}e^{ik^s_mx}
      +b_{m}e^{-ik^s_m x})F_{n-m}e^{-in\omega t},\\
      &(& x_0\leq x\leq x_0+\delta)\ \\
\Psi_r &=& \sum\limits_{n=-\infty}^{\infty} (B_n^i
      e^{-ik_n x}+B_n^o e^{ik_n x})e^{-iE_n t/\hbar},\ (x > x_0+\delta) 
\end{eqnarray*}
where $A_n^i$ and $B_n^i$ are the probability amplitudes of
the incoming waves from
the left and right, respectively, while
$A_n^o$ and $B_n^o$ are those of the outgoing waves.
We can characterize the barrier by a scattering matrix $S$
which is a matrix connecting the incoming
and outgoing channels

$$
\left (
\begin{array}{c}
{\bf A}^o \\
{\bf B}^o
\end{array} \right )
=S\left (
\begin{array}{c}
{\bf A}^i \\
{\bf B}^i
\end{array} \right ),
$$
where the $S$ matrix can be derived by the
matching conditions for the wave function $\Psi(x,t)$
and its derivative $\partial_x \Psi(x,t)$ at
$x=x_0$ and $x=x_0+\delta$.
After eliminating $a_{m}$ and $b_{m}$,
we have~\cite{Burmeister}
\begin{equation}
\label{S}
S=\left (
\begin{array}{cc}
R_{\rightarrow} & T_{\leftarrow} \\
T_{\rightarrow} & R_{\leftarrow}
\end{array} \right ),
\end{equation}
where $T_{\leftarrow} =  L_L T L_R^{-1},$ $R_{\leftarrow}=
L_R^{-1} {R} L_R^{-1},$
$T_{\rightarrow} = L_R^{-1} T L_L,$ and $ R_{\rightarrow}
=L_L R L_L$.
Here the left (right) arrow indicates incoming waves from right (left),
the matrices $L_L$ and $L_R$
are defined as $(L_L)_{mn}=exp[ik_n x_0]\delta_{mn}$ and
$(L_R)_{mn}=exp[ik_n(x_0+\delta)]\delta_{mn}$.
$T$ and $R$ are given by
\begin{eqnarray}
\label{T-tilde}
T &=& (C_1^{-1}D_1+C_2^{-1}D_2)/2, \\
\label{R-tilde}
R &=& (C_1^{-1}D_1-C_2^{-1}D_2)/2,
\end{eqnarray}
where
\begin{eqnarray*}
C_1 &=& (L_s-\widetilde{I})K_s F^{\dagger}
      -(L_s+\widetilde{I})F^{\dagger} K, \\
D_1 &=& -(L_s-\widetilde{I})K_s F^{\dagger}
      -(L_s+\widetilde{I})F^{\dagger} K, \\
C_2 &=& (L_s+\widetilde{I})K_s F^{\dagger}
      - (L_s-\widetilde{I})F^{\dagger} K, \\
D_2 &=& (L_s+\widetilde{I})K_s F^{\dagger}
      +(L_s-\widetilde{I})F^{\dagger} K,
\end{eqnarray*}
with the matrices $(L_s)_{mn}=exp[ik^s_n\delta]\delta_{mn}$,
$(K_s)_{mn}=k^s_n\delta_{mn}$,
$K_{mn}=k_n\delta_{mn}$,
$\widetilde{I}$ as the unit matrix
and $F^{\dagger}$
as the Hermitian conjugate of $F$.
The electronic transport properties of the
scattering region may be obtained straightforward from
Eq.(\ref{S}).

The above method may be generalized to $l$ time-periodic barriers
described by
\begin{equation}
\label{potential3}
V(x,t)=\left\{ \begin{array}{cc}
0, & x<0,\ x>a_l  \\
V_1(x,t), & 0 \leq x <a_1 \\
V_2(x,t), & a_1 \leq x <a_2 \\
\vdots  & \vdots \\
V_l(x,t), & a_{l-1} \leq x \leq  a_l
\end{array}
\right.
\end{equation}
where $V_1(x,t)=V_1^{0}+V_1 cos(\omega_1 t+\phi_1)$,
$V_2(x,t)=V^0_2+V_2 cos(\omega_2 t+\phi_2),\cdots$
and $V_l(x,t)=V^0_l+V_l cos(\omega_l t+\phi_l)$.
This potential may be more a realistic model for experiments.
Obviously the transport properties for each barrier
can be characterized by
an $S$ matrix given by
$$
S^\alpha=\left (
\begin{array}{cc}
R_{\rightarrow}^\alpha & T_{\leftarrow}^\alpha \\
T_{\rightarrow}^\alpha & R_{\leftarrow}^\alpha
\end{array} \right ),
$$
where $\alpha=1,2,\cdots,l$,
$T_{\rightarrow}^\alpha$, $T_{\leftarrow}^\alpha$,
$R_{\rightarrow}^\alpha$ and $R_{\leftarrow}^\alpha$
can be derived by the same method presented above.
Now the propagating mode $E_n$ should be replaced by
\begin{equation}
\label{energy}
E(n_j)=E_0+\sum\limits_{n_j=-\infty}^\infty
n_j\hbar\omega_j.\ \ (j=1,\cdots,l)
\end{equation}
The associated transfer-matrix $M^\alpha$ for
the $\alpha$th barrier
may be derived directly from the $S^\alpha$ matrix
$$
M^\alpha=\left (
\begin{array}{cc}
(T_{\leftarrow}^\alpha)^{-1}
& -(T_{\leftarrow}^\alpha)^{-1}R_{\rightarrow}^\alpha \\
R_{\leftarrow}^\alpha (T_{\leftarrow}^\alpha)^{-1}
& T_{\rightarrow}^\alpha-R_{\leftarrow}^\alpha
(T_{\leftarrow}^\alpha)^{-1}R_{\rightarrow}^\alpha
\end{array} \right ).
$$
The total transfer-matrix $M^{t}$ for all those barriers
is determined by
$$
M^{t}
=\left (
\begin{array}{cc}
M_{11}^t & M_{12}^t \\
M_{21}^t & M_{22}^t
\end{array} \right )
=M^l M^{n-l} \cdots M^1,
$$
where $M^t_{ij}\ (i,j=1,2)$ are
the partitioned matrices with the same size
as $T^\alpha_{\rightarrow}$.
The total scattering matrix $S^{t}$ can be derived from $M^t$ as
$$
S^t=
\left (
\begin{array}{cc}
-(M_{11}^t)^{-1}M_{12}^t & (M_{11}^t)^{-1} \\
M_{22}^t-M_{21}^t(M_{11}^t)^{-1}M_{12}^t & M_{21}^t (M_{11}^t)^{-1}
\end{array} \right ).
$$

In each cycle a net charge current may
pass through the scattering region in the direction determined from
the detailed form of $S^t$ matrix.
We define a net transmission coefficient (for an incoming wave in mode
$E_0=E$) by
$$
T_{net}=\sum\limits_{E(n_j)>0}\sqrt{\frac{2E(n_j)}{m^*}}
(|T^t_{\rightarrow ,n_j0}(E_0)|^2
-|T^t_{\leftarrow ,n_j0}(E_0)|^2).
$$
The average net current per period $\tau$ (for $E_0$) through
those barriers is $j(E_0)=T_{net}(E_0)$.
If the system is connected through two ideal leads
to two electron reservoirs with the same
chemical potential $\mu$, the average pumped current
per period $\tau$ is given by~\cite{Burmeister,Landauer}
\begin{equation}
\label{current}
I(\mu)=e\int_0^{\infty}dE g(E)f(E-\mu)T_{net}(E),
\end{equation}
where $g(E)=\sqrt{2m^*/E}/h$ is the density of electrons contributing to the
current in one direction, and $f(E-\mu)$ is the Fermi-Dirac distribution.
At zero temperature, it becomes
\begin{equation}
\label{I_mu}
I(\mu)=\frac{2e}{h}\int_0^{\mu}dE \sqrt{\frac{m^*}{2E}}
T_{net}(E).
\end{equation}

Another important quantity is the Wigner delay time which gives
the time delay of the scattered electron due
to its interaction with the scattering field (
here the oscillating potential). It relates to
the S-matrix by~\cite{Wigner-smith}
\begin{equation}
\label{time}
\tau_w(E)=-\frac{i\hbar}{N_c}Tr[(S^t)^\dagger\frac{d S^t}{dE}]
=-\frac{i\hbar}{N_c}\frac{d}{dE}ln(detS^t),
\end{equation}
where $N_c$ is the number of open channels.
Physically, the Wigner
time represents the time spent by a wave packet passing through the
scattering region.
The charge pumping is supposed to be adiabatic
when $\tau$ is much greater than the Wigner delay time $\tau_w$.

It is obvious that the net charge transfer in one cycle is zero
for a single time-periodic barrier since
$T_{\rightarrow}=T_{\leftarrow}$.
Then the simplest system which may induce
the nontrivial charge pumping should include at least
two barriers.
As an example we consider a mesoscopic system with two
time-periodic
barriers connected through ideal leads to
two electron reservoirs with
the same chemical potentials $\mu$.
The potentials are described by $V_1(x,t)=V_1^0+V_1 cos(\omega t)$,
$V_2(x,t)=V_2^0$,
and $V_3(x,t)=V_3^0+V_3cos(\omega t+\phi)$.
This appears to be a simplified model for
the Switkes {\sl et al}'s experiment, nevertheless it turns out that
some essential characteristics can be exhibited, as we will address below.
In the following numerical calculations, $\mu =75\ mev$,
$m^*=0.067m_e$ (with $m_e$ as the mass of the free electron), 
$V_2^0=-30\ mev$, $V_2=0$, and
$N_c$ is determined by a natural condition:
$|T_{\rightleftharpoons}|^2+|R_{\rightleftharpoons}|^2-1.0\leq c_e$
with $c_e$ ($=1.0\times 10^{-4}$ in this paper)
as a defined error.

\begin{figure}
\label{fig1}
\epsfxsize=7.5cm
\epsfbox{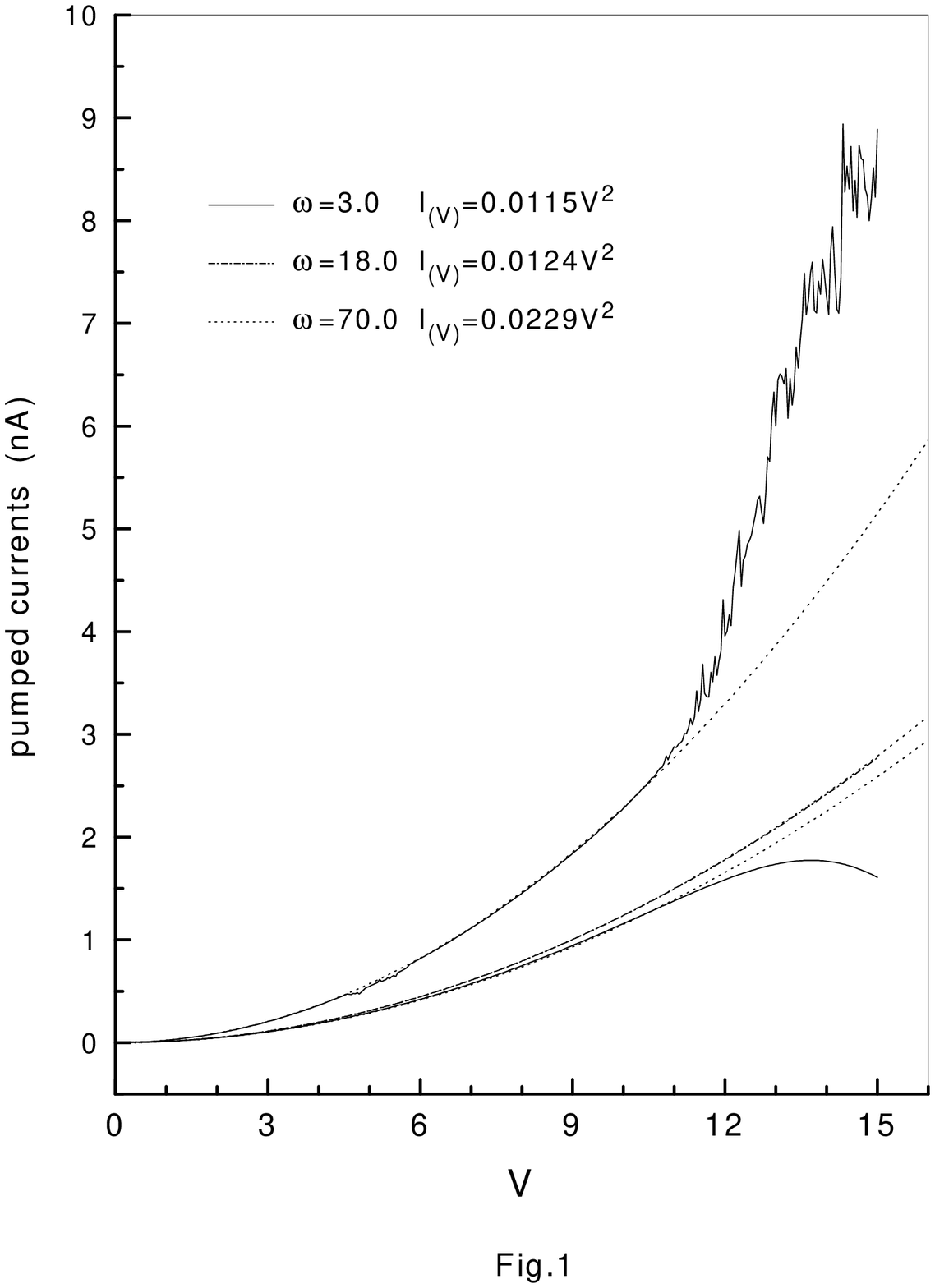}
\caption{The pumped currents $I(\phi=\pi/2)$
versus the barrier height $V$ for different
pumping frequencies. Dotted lines 
fit the relation $I_{(V)}\propto V^2$.}
\end{figure}

The general characteristics of quantum pumping are
shown in Figs.1 and 2.
The parameters in Figs.1, 2, and 3 are chosen as
$V^0_1=V^0_3=50\ mev$, 
and $V_1=V_3=V$.
Figure 1 shows that the pumped current $I(V)$
is proportional to $V^2$ for small pumping amplitude $V$,
with the proportional factor depending on
the driving frequency  ($\hbar=1$).
But it deviates from
$V^2$-dependence for the strong pumping case.
On the other hand, the pumped current is sinusoidal dependence
on $\phi$ for weak pumping, and becomes
nonsinusoidal dependence on $\phi$ when $V$ increases,
as seen in Fig.2.
Another important characteristic shown in Fig.2
is that $I(\pi+\phi,V)=-I(\phi,V)$
for all amplitude strengths, and $|I(\phi,V)|$
is maximum at $\phi=\pi/2$ or $\phi=3\pi/2$ for weak pumping.
Remarkably, these results for $I(\phi,V)$ are in
agreement with the experimental observation in Ref.[3].

\begin{figure}
\label{pumpfig2}
\epsfxsize=7.5cm
\epsfbox{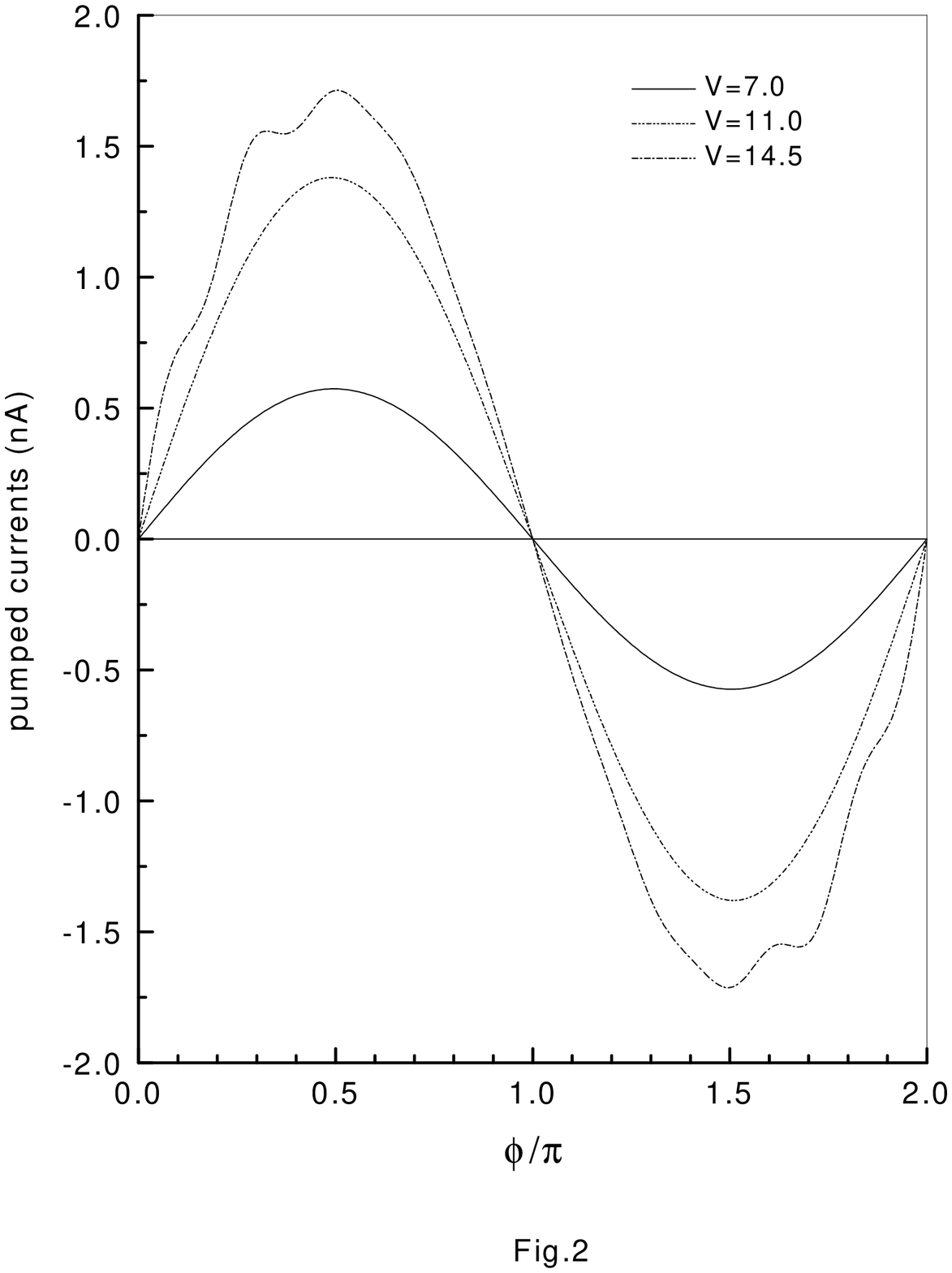}
\caption{The pumped currents versus the phase difference $\phi$ for
three different $V$ and $\omega=3.0$.}
\end{figure}

Figure 3 shows that sharp peaks in the Wigner time occur 
at the resonance insert energies $E_0=n\hbar \omega$.
Besides, jumps in the pumped current
as a function of $E_0$ appear at the peak positions
of the Wigner time.
The direction of the current depends crucially on
the insert energy.
It is interesting to note that the adiabatic condition
is not necessary in our calculations.
Figure 3 indicates that the maximum value of
$\tau_w$ is about $5.5\ ns$ for $\hbar\omega=6.0\ mev$ (corresponding to
$\tau\sim 0.7\ ps$ ), which is much greater than
the pumping cyclic time $\tau$. Then we may say that
the method described here is beyond adiabaticity.
Actually,
the nonadiabatic effects are only important for
the strongly photon-assisted transport
since $\tau_w$ is greater than
$\tau$ only if the energy of the incoming wave
is approximately equal to
the resonance energy for photon-assisted tunneling.
Physically,
by emitting or absorbing photons, the outgoing
waves may be at the quantum states different  from that of the incoming wave.
Consequently, the adiabatic condition, which requires that
the quantum state is at the same instant state in
the whole evolution, is not satisfied.
Note that the formula derived by
Brouwer\cite{Brouwer} may be valid merely under the adiabatic condition,
and thus the method developed here may be quite useful.

\begin{figure}
\label{pumpfig3}
\epsfxsize=7.5cm
\epsfbox{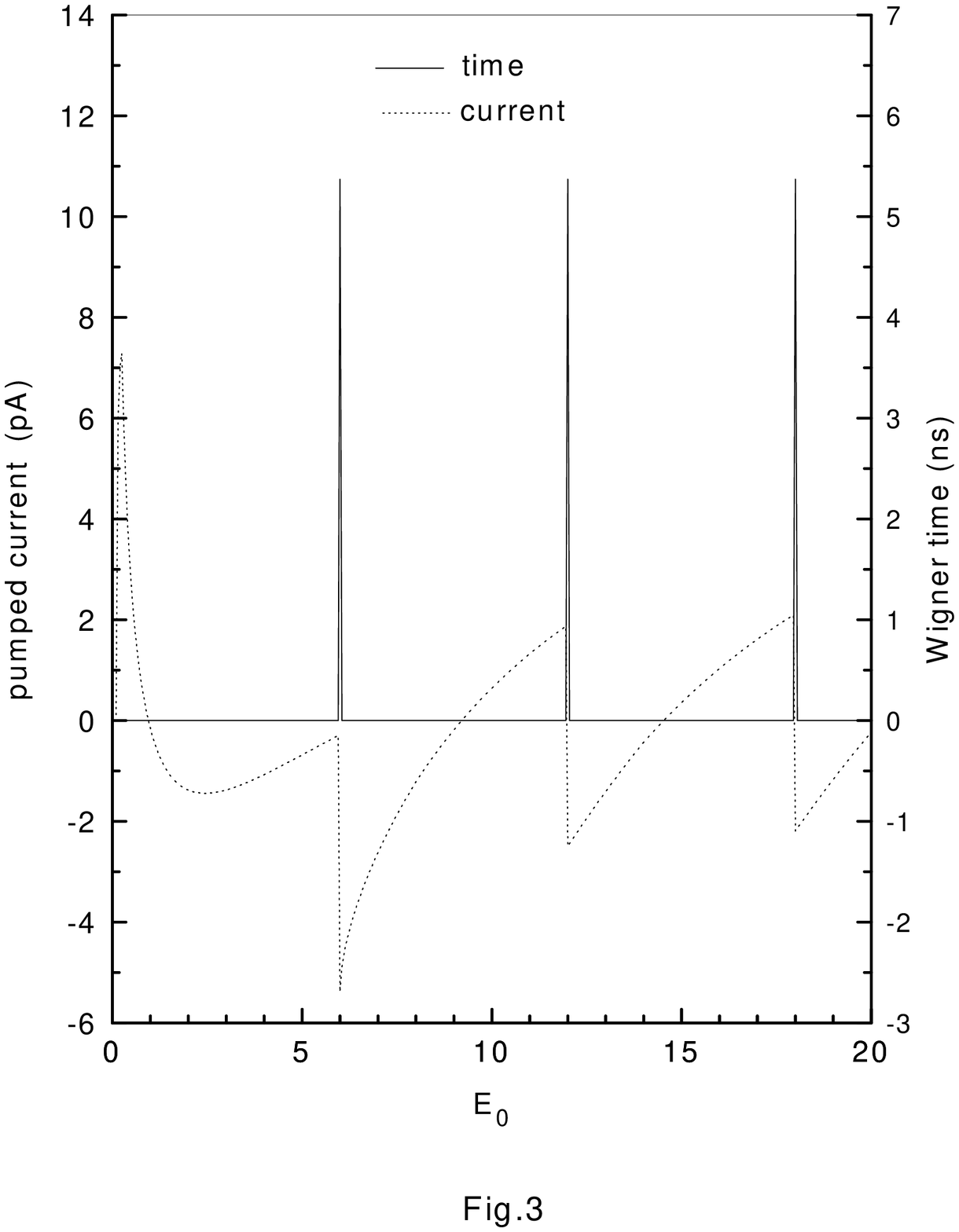}
\caption{The pumped current and the Wigner delay time versus
the insert energy $E_0$ for $V=7.0$ and $\omega=6.0$.}
\end{figure}

It is quilt intriguing to note from Fig.4 that
$I(\phi=0)$  is nonzero for $V_1\not=V_3$,
while the corresponding areas enclosed by
the path in the parameter space $\{ V_1(x,t),V_3(x,t) \}$
are zero. Although the pumped currents in the above case were predicated 
to be zero under the adiabatic approximation, the deviation from zero 
is reported experimentally at strong pumping,\cite{Switkes}
just as we observed here in terms of a rigorous theoretical analysis
which is also valid for strong pumping.
Moreover,  from comparison with that  $I(\phi=0)=0$ for $V_1=V_3$, 
it is now clear that the present nonzero pumped currents 
stem from the spatial asymmetry 
of potentials $V_1\not=V_3$,
which is coincident with the result obtained by Wagner
in Ref.\cite{Wagner}:
the nonzero currents may be observed in a single osscillating
potential but with asymmetric static potential.
Actually, to observe a pump current at zero applied
bias,  it  seems that the inversion symmetry should be broken,
either in real or in $k$ space.

\begin{figure}
\label{pumpfig4}
\epsfxsize=7.5cm
\epsfbox{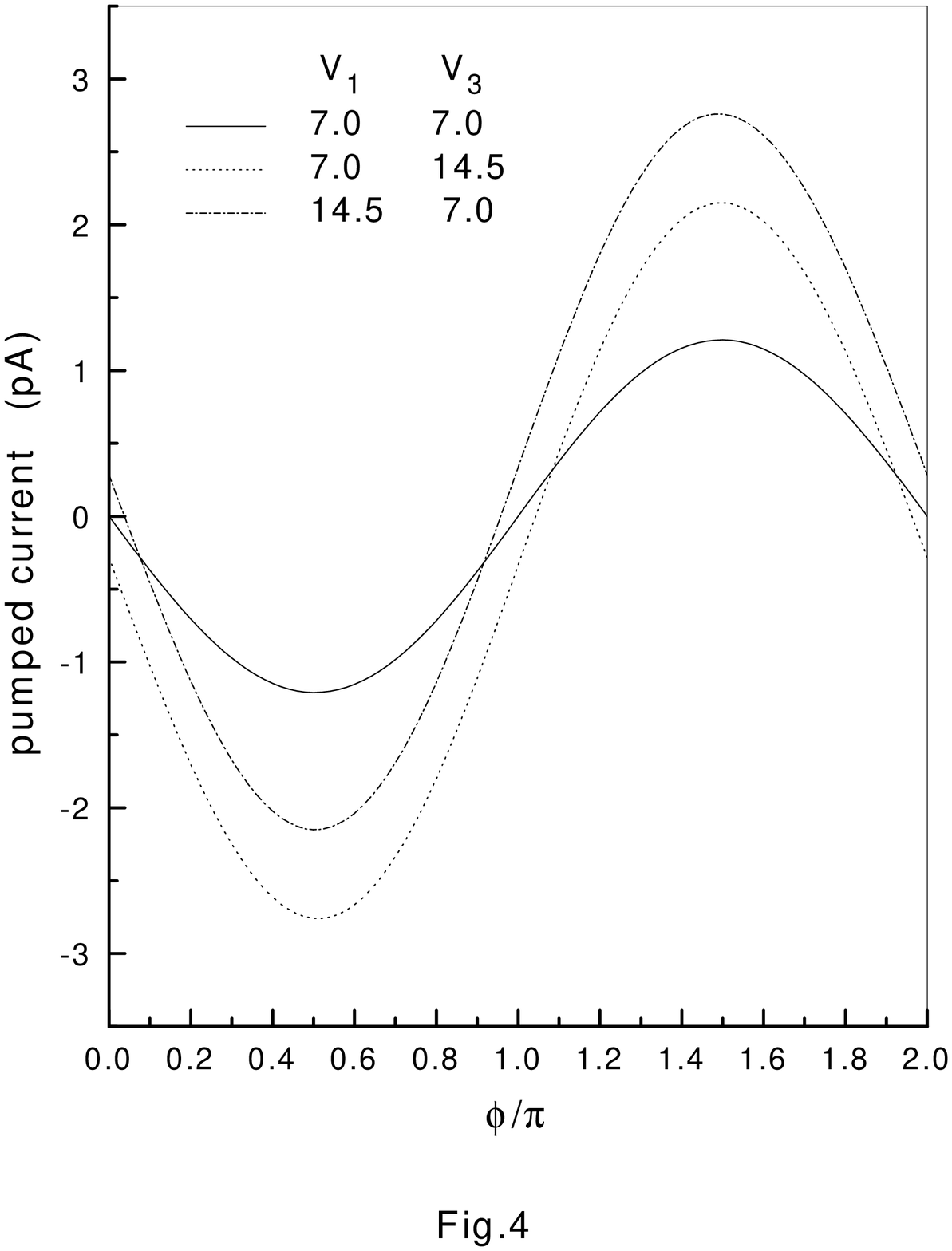}
\caption{The pumped currents versus the phase difference $\phi$ 
for $\omega=3.0$ for different $V_1$ and $V_3$.}
\end{figure}

The fact that $I(\phi=0)$ is nonzero at strong pumping
may be understood based on a scenario of the nonadiabatic
geometric phase.\cite{Zhu}
Pumped currents are determined by geometric phase accumulated in
the evolution.~\cite{Thouless,Switkes,Altshuler} Under
the adiabatic approximation, $I(\phi=0)=0$
is predicted theoretically because the corresponding adiabatic 
geometric phase is zero. 
 While it is now clear that  the nonadiabatic 
geometric phase may be nonzero 
even in the case where the area enclosed by the path in the
parameter space is zero
(thus the adiabatic phase is zero).\cite{Zhu}
Therefore, the nonadiabatic correction
to the currents should be taken into account for strong pumping
whenever the adiabatic condition is not well satisfied.
Physically,
it is reasonable to believe that the observed nonzero
pumping at phase $\phi=0$
for the strong pumping stems from the nonadiabatic correction
when the inversion symmetry is  broken.
Practically, the asymmetric spatial potential might be present 
in the experiment,  which may originate from either the 
shape-distorting ac voltages, or from the internal potential
established during transport.\cite{Buttiker} Since
the current calculated in this approach is conserved
since $|T_{\rightleftharpoons}|^2+|R_{\rightleftharpoons}|^2=1.0$, 
 no internal potential appears explicitly in
the present formulism.
It is
worth pointing out that nonzero pumped currents for $\phi=\pi$ are also
seen in Fig.4, which seems to contradict with that
 in Ref \cite{Switkes}.
Also note that a  nonzero $I(\phi=\pi)$
was also predicted by another totally different theoretical
study\cite{Wang}, so this contradication is still
an interesting open question at present.

In summary, we developed a method to calculate the pumped current and
Wigner delay time in a mesoscopic system with
a series of time-periodic barriers connected to two
electron reservoirs, which appears to be applicable for 
strong pumping cases.

We thank the support from a RGC grant of
Hong Kong(Grant No. HKU7118/00P) and a CRCG grant
at the University of Hong Kong.



\begin{thebibliography}{999}


\bibitem{Thouless} D. J. Thouless, Phys. Rev. B {\bf 27}, 6083 (1983).
\bibitem{Zhou} F. Zhou, B. Spivak, and B. Altshuler,
Phys. Rev. Lett. {\bf 82}, 608 (1999).
\bibitem{Brouwer} P. W. Brouwer, Phys. Rev. B {\bf 58},
R10135 (1998).
\bibitem{Niu} Q. Niu, Phys. Rev. Lett. {\bf 64}, 1812 (1990).
\bibitem{Switkes} M. Switkes, C. M. Marcus, K. Campman, and A. C.
Gossard, Science, {\bf 283}, 1905 (1999).
\bibitem{Altshuler} B. L. Altshuler and L. I. Glazman, Science,
{\bf 283}, 1864 (1999).
\bibitem{Buttiker} M. B\"{u}ttiker, H. Thomas, and A. Pr\^{e}tre,
Z. Phys. B {\bf 94}, 133 (1994).
\bibitem{Andreev} A. Andreev and A. Kamenev,
cond-mat/0001460.
\bibitem{Aleiner} I. L. Aleiner and A. V. Andreev,
Phys. Rev. Lett. {\bf 81}, 1286 (1998);
T. A. Shutenko, I. L. Aleiner, and B. L. Altshuler,
Phys. Rev. B {\bf 61}, 10 366 (2000).
\bibitem{Levinson} Y. Levinson, O. Entin-Wohlman, and P. W\"{o}lfle,
Phys. Rev. Lett. {\bf 85}, 634 (2000).
\bibitem{Wagner} M. Wagner, Phys. Rev. Lett. {\bf 85}, 174 (2000);
M. Wagner and F. Sols, {\sl ibid}. {\bf 83}, 4377 (1999);
M. Wagner, {\sl ibid.} {\bf 76}, 4010 (1996).
\bibitem{Wei} Y. Wei, J. Wang, and H. Guo, Phys. Rev. B {\bf 62}, 9947 (2000).
\bibitem{Burmeister} G. Burmeister and K. Maschke, Phys. Rev.
B {\bf 57}, 13 050 (1998); W. Li and L. E. Reichl,
{\sl ibid.} {\bf 60}, 15 732 (1999).
\bibitem{Landauer} R. Landauer, Philos. Mag. {\bf 21}, 863 (1970);
M. B\"{u}ttiker, Phys. Rev. Lett. {\bf 57}, 1761 (1986).
\bibitem{Wigner-smith} E. P. Wigner, Phys. Rev. {\bf 98}, 145 (1955);
F. T. Smith, {\sl ibid.} {\bf 118}, 349 (1960).
\bibitem{Zhu} S. L. Zhu and Z. D. Wang, Phys. Rev. Lett. {\bf 85}, 1076 (2000);
S. L. Zhu, Z. D. Wang, and Y. D. Zhang, Phys. Rev. B {\bf 61}, 1142 (2000).
\bibitem{Wang} B. Wang, J. Wang, and H. Guo, Phys. Rev. {\bf B},
073306 (2002).




\end{thebibliography}
\end{document}